\documentclass[final,5p,times,sort,compress]{elsarticle}
\usepackage{lmodern}
\usepackage[normalem]{ulem}

\usepackage{amssymb}
\usepackage{amsmath}
\usepackage[colorlinks=true, linkcolor=blue, citecolor=blue, urlcolor=blue]{hyperref}
\usepackage[nameinlink]{cleveref}
\crefname{figure}{Figure}{Figures}
\Crefname{figure}{Figure}{Figures}
\crefname{equation}{Eq.}{Eqs.}
\Crefname{equation}{Eq.}{Eqs.}

\usepackage{booktabs}

\usepackage{verbatim}

\usepackage{microtype}

\usepackage[htt]{hyphenat}

\usepackage{soul}

\usepackage{xcolor}
\usepackage{listings}

\definecolor{darkgreen}{rgb}{0,0.45,0}
\definecolor{myblue}{rgb}{0,0,0.8}
\definecolor{myturquoise}{rgb}{0,0.5,0.5}

\lstdefinestyle{mypython}{
  language=Python,
  basicstyle=\ttfamily\small,
  keywordstyle=\color{black},
  morekeywords=[1]{import,as},
  keywordstyle=[1]\color{darkgreen},
  morekeywords=[2]{multisphere,mss},
  keywordstyle=[2]\color{myblue},
  commentstyle=\color{myturquoise},
  showstringspaces=false,
  breaklines=true
}

\usepackage{tabularx}

\usepackage{cuted}
\usepackage{capt-of}
\journal{SoftwareX}

\begin{document}
\renewcommand{\labelenumii}{\arabic{enumi}.\arabic{enumii}}

\begin{frontmatter}
 


\title{\texttt{multisphere}: a Python implementation of the \underline{M}ulti \underline{S}phere \underline{S}hape generator (MSS) for DEM simulations}

\author[label1]{F. Buchele}
\author[label1]{P. Müller}
\author[label1]{A. Moradian}
\author[label1]{T. Pöschel}
\address[label1]{Institute for Multiscale Simulation, Friedrich-Alexander-University Erlangen-Nürnberg, Germany}

\begin{abstract}
\textit{\texttt{multisphere} is an open-source Python package for generating multi-sphere representations of complex particles for use in DEM simulations. It reconstructs triangulated surface meshes and voxelized volumes as sets of intersecting spheres and provides tools for evaluation, visualization, and export.}
%
\end{abstract}

\end{frontmatter}
\vspace{-0.8\baselineskip}
\begin{strip}
\vspace{-1.0\baselineskip}
\section*{Metadata}
\vspace{-0.5\baselineskip}
\centering
\captionof{table}{Code metadata}
\label{codeMetadata}
\vspace{0.3\baselineskip}

\begin{tabularx}{\textwidth}{@{}p{0.08\textwidth} p{0.42\textwidth} X@{}}
\toprule
\textbf{Nr.} & \textbf{Code metadata description} & \textbf{Metadata} \\
\midrule
C1 & Current code version & v1.1.0 \\
C2 & Permanent link to code/repository used for this code version &
\url{https://github.com/FelixBuchele/multisphere} \\
C3 & Permanent link to Reproducible Capsule & None \\
C4 & Legal Code License & GNU GPL 3.0 \\
C5 & Code versioning system used & git \\
C6 & Software code languages, tools, and services used & python $\geq$ 3.10 \\
C7 & Compilation requirements, operating environments \& dependencies &
numpy, trimesh, scipy, scikit-image, matplotlib, pyvista, manifold3D \\
C8 & If available Link to developer documentation/manual &
\url{https://github.com/FelixBuchele/multisphere/blob/main/README.md} \\
C9 & Support email for questions &
\href{mailto:felix.buchele@fau.de}{felix.buchele@fau.de} \\
\bottomrule
\end{tabularx}
\end{strip}

\section{Motivation and significance}
The discrete element method (DEM) is a numerical technique for the simulation of granular and particulate systems in fields such as materials science, geomechanics, and process engineering \cite{Cundall.1979, Poeschel.2005, Matuttis.2014}. In many cases, the macroscopic behavior of such systems depends strongly on the shapes of individual particles, which influence packing structure, contact interactions, and flow dynamics \cite{Lu.2015, Zhao.2023}. 

The multi-sphere approach is a widely used model for non-spherical particles, whose shape $\mathcal{S}$ is represented by a set of $n$ spheres $\tilde{\mathcal{S}}_i$. Their radii $R_i$ and positions $\vec{r}_i$. must be chosen such that
\begin{equation}
   \mathcal{\tilde{S}} \equiv \bigcup_{i=0}^{n-1} \mathcal{\tilde{S}}_i(R_i, \vec{r}_i)\,, 
   \label{eq:sphereRep}
\end{equation}
approximates $\mathcal{S}$ as closely as possible for given $n$.

\begin{figure}
    \centering
    \includegraphics[width=1.0\linewidth]{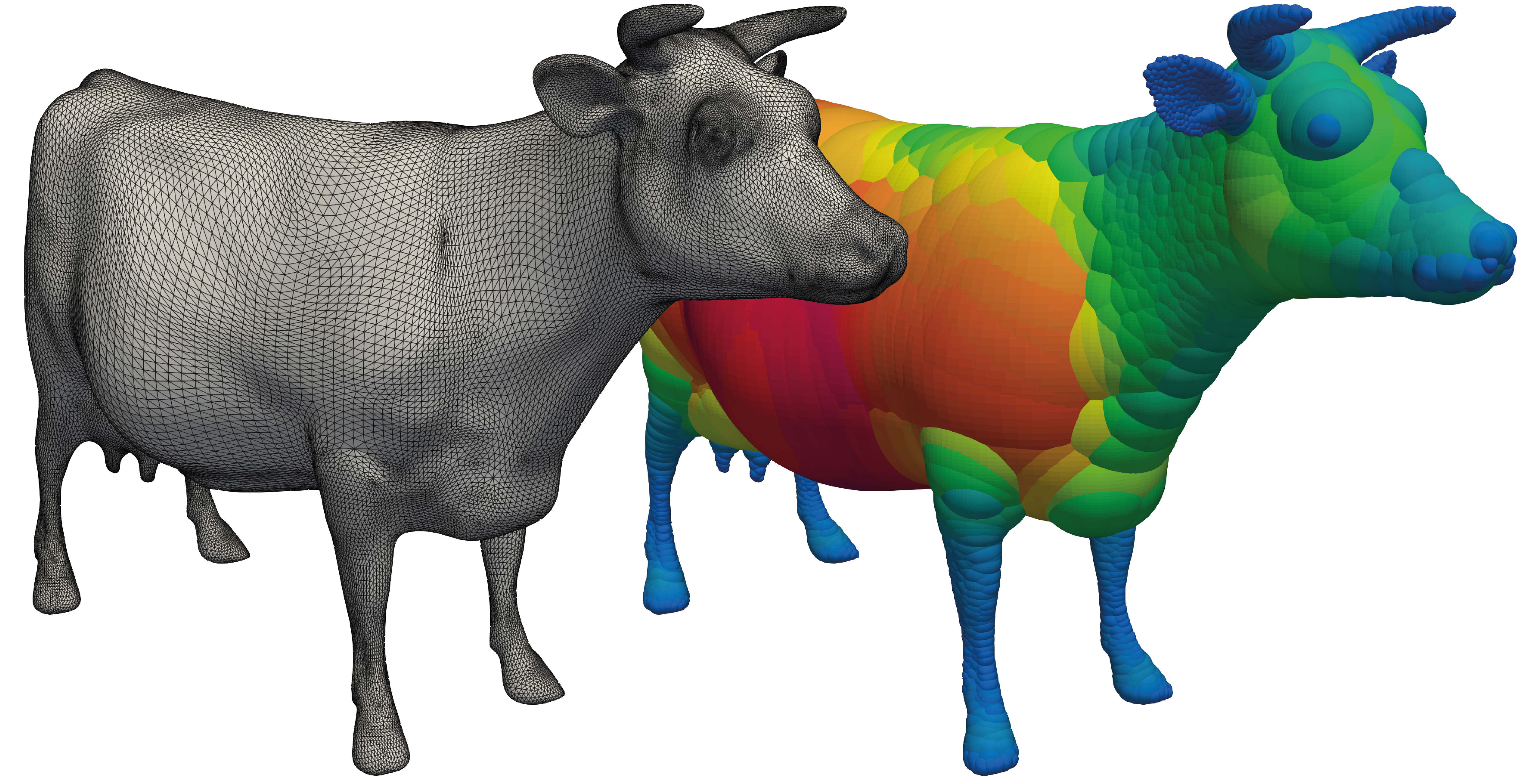}
    \caption{Mesh model of a cow \cite{Cow2:Viewpoint} and its multisphere representation with $n=2000$. Color indicates sphere sizes.
    }
    \label{fig:cowExample}
\end{figure}

A key advantage of the multi-sphere representation is that contact detection and the evaluation of interaction forces for complex-shaped particles reduce to sphere–sphere interactions \cite{Poeschel.1993, Buchholtz.1994, Buchholtz.1996, Hubbard.1996, Favier.1999}. An example of a multisphere representation of a complex shape is given in \cref{fig:cowExample}. 

Determining the set $\left\{\left(R_i,\vec{r}_i\right)\right\}, i=0,\dots,n-1$ is an optimization problem in which the mismatch 
\begin{equation}
%
\left| \mathcal{S}\triangle \tilde{\mathcal{S}} \right|    \equiv 
\left|\left( \mathcal{S} \cup \tilde{\mathcal{S}} \right) 
\setminus \left({\mathcal S}\cap \tilde{\mathcal{S}}\right) 
\right|    \,.
    \label{eq:mismatch-def}
\end{equation}
between the target shape $\mathcal{S}$ and its multi-sphere approximation $\tilde{\mathcal{S}}$ is minimized. Besides the number of spheres $n$, additional constraints may be imposed, such as the minimum sphere size and the minimum distance between spheres.
This optimization is a challenging problem in computational geometry, and several multi-sphere particle generators have been proposed \cite{Angelidakis.2021, Canbolat.2025, Favier.1999, Amberger.2012, Ferellec.2010}.

The \underline{M}ulti-\underline{S}phere \underline{S}hape generator (MSS) algorithm was introduced as a theoretical concept in \cite{Buchele.2026}. MSS requires fewer spheres than other particle generators to achieve a given mismatch value. Unlike other particle generators, MSS preserves symmetry properties, such that the generated particle model $\tilde{\mathcal{S}}$ exhibits the same symmetries as the target shape ${\mathcal{S}}$.

MSS is computationally efficient \cite{Buchele.2026}, making it well suited for applications that require frequent particle reconstruction, such as simulations involving particle fracture and plastic deformation.

This article presents a Python implementation of MSS. The underlying mathematical and algorithmic concepts are described in \cite{Buchele.2026}.

\section{Software description}

\texttt{multisphere} is implemented in Python and requires Python version $\geq$~3.10. Core functionality depends on \texttt{numpy}, \texttt{scipy}, \texttt{scikit-image}, and \texttt{trimesh}. Optional dependencies are used for visualization (\texttt{pyvista}) and for mesh-based boolean operations during mesh generation (\texttt{manifold3D}).

The package is available on the Python Package Index (PyPI) and can be installed with \texttt{pip}. \texttt{pip install multisphere} installs the core package, and \texttt{pip install multisphere[full]} installs the core package together with all optional features. The source code is available in a public GitHub repository (see Table \ref{codeMetadata}).

\subsection{Software architecture}

\begin{figure*}[!t]
    \centering
    \includegraphics[width=0.75\linewidth]{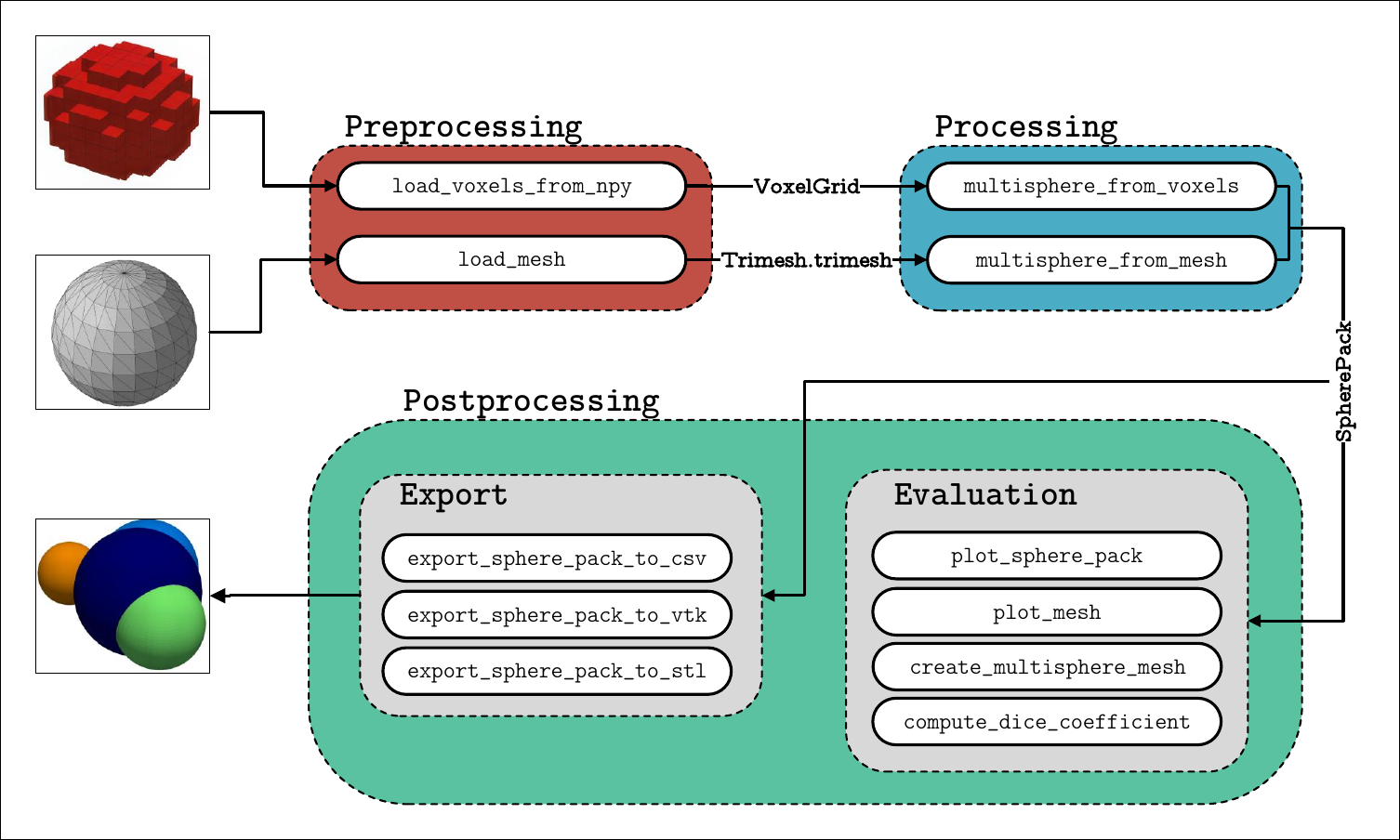}
    \caption{High-level architecture of the \texttt{multisphere} package. There are three main stages of the workflow involving public functions.  
    }
    \label{fig:mss-architecture}
\end{figure*}

\texttt{multisphere} follows a three-stage workflow:
preprocessing, processing, and postprocessing. Target shapes are provided either as surface meshes or voxelized volumes. Reconstruction is performed on voxel data, so mesh input is voxelized internally when required. The result is stored as a \texttt{SpherePack}, which can then be evaluated, visualized, or exported. \Cref{fig:mss-architecture} summarizes the software architecture and public interface. Details on the MSS algorithm can be found in \cite{Buchele.2026}.

\subsection{Software functionalities}

\subsubsection*{Core data structures}
\label{sec:data_types}

\texttt{multisphere} uses two core data  structures to connect the different stages of the workflow: \texttt{VoxelGrid} and \texttt{SpherePack}.

\texttt{VoxelGrid} is a target particle as a three-dimensional voxel grid. It is used both for voxelized input data and for intermediate voxel-based distance fields required during reconstruction. \texttt{VoxelGrid} stores the voxel size and the spatial origin, allowing the discrete grid to be mapped to physical coordinates.

\texttt{SpherePack} stores the positions and radii of the spheres consituting the multisphere particle model, see \cref{eq:sphereRep}. All quantities in \texttt{SpherePack} are expressed in physical units.

\subsubsection*{Public interface overview}

\Cref{tab:public_interface} lists the main public functions of \texttt{multisphere}, and \cref{fig:code_example} shows an example workflow.

\begin{table}[!t]
\scriptsize
\centering
\caption{Public interface of the \texttt{multisphere} package.}
\label{tab:public_interface}
\begin{tabularx}{\columnwidth}{@{}>{\raggedright\arraybackslash}p{0.36\columnwidth}
                                  >{\centering\arraybackslash}p{0.07\columnwidth}
                                  >{\raggedright\arraybackslash}X@{}}
\toprule
\textbf{Function} & \textbf{Stage} & \textbf{Description} \\
\midrule
\texttt{load\_mesh} & 1 & Load a triangulated surface mesh from a mesh file. \\
\texttt{load\_voxels\_from\_npy} & 1 & Load a voxelized target shape stored as a NumPy array. \\
\texttt{multisphere\_from\_voxels} & 2 & Generate a multi-sphere representation from voxelized target shape. \\
\texttt{multisphere\_from\_mesh} & 2 & Generate a multi-sphere representation from a mesh target shape via voxelization. \\
\texttt{create\_multisphere\_mesh} & 3 & Construct a surface mesh representing the union of spheres in a \texttt{SpherePack}. \\
\texttt{compute\_dice\_coefficient} & 3 & Compute the Dice similarity coefficient between two meshes. \\
\texttt{plot\_mesh} & 3 & Interactive visualization of surface meshes. \\
\texttt{plot\_sphere\_pack} & 3 & Interactive visualization of a multi-sphere representation. \\
\bottomrule
\end{tabularx}
\end{table}

\subsubsection*{Preprocessing}

Meshes stored in STL, OBJ or PLY format are loaded using \texttt{load\_mesh}, which returns a \texttt{trimesh.Trimesh} object. Voxelized particle target shapes stored as NumPy arrays are loaded using \texttt{load\_voxels\_from\_npy}, which returns a \texttt{VoxelGrid}. For voxel inputs, the voxel size and origin can be specified to map the grid to physical coordinates.

\subsubsection*{Processing}

\texttt{multisphere} can be applied   either to voxelized target shape descriptions using \texttt{multisphere\_from\_voxels} or to surface meshes using \texttt{multisphere\_from\_mesh}. In the latter case, the input mesh is first voxelized, with the voxel size 
\begin{equation}
    \Delta = \frac{L_{\min}}{\texttt{div}},
\end{equation}
derived from the smallest edge of the axis-aligned bounding box $L_{\min}$ and the specified number \texttt{div} of voxels along this edge. Both functions return a \texttt{SpherePack}.

The reconstruction process is controlled by three main termination criteria: a target reconstruction accuracy, a minimum admissible sphere radius, and a maximum number of spheres. At least one of these criteria must be specified. 

The reconstruction accuracy is expressed through the Dice-S{\o}rensen coefficient \cite{Dice.1945} (as cited in \cite{Levy.2025}), 
\begin{equation}
    D \equiv \frac{2 \sum\limits_{i,j,k} \left( S_{ijk} \cdot \tilde{S}_{ijk} \right) }{\sum\limits_{i,j,k} S_{ijk} + \sum\limits_{i,j,k} \tilde{S}_{ijk} }\,,
    \label{eq:Dice}
\end{equation}
quantifying the similarity of the target particle shape $\mathcal{S}$ and its multi-sphere approximation $\tilde{\mathcal{S}}$.

For mesh-based input, voxelization introduces discretization errors, therefore, the reconstructed particle model, $\tilde{\mathcal{S}}$ may locally extend beyond the target surface mesh. This property can lead to problems in certain applications, for example when the fragments of a granular particle that has just broken are both modeled as multi-sphere particles. If the multi-sphere model of one or both fragments extends beyond the fracture surface, this could result in the DEM behaving in an unphysical manner. 

The optional flag \texttt{confine\_mesh} prevents the extension of $\tilde{\mathcal{S}}$ beyond the target mesh by reducing the sphere radii accordingly.

\subsubsection*{Postprocessing}

The function \texttt{create\_multisphere\_mesh} converts the multi-sphere approximation to a \texttt{trimesh.Trimesh} surface mesh. This mesh can be used for visualization and for mesh-based accuracy evaluation with \texttt{compute\_dice\_coefficient}, which computes the Dice similarity between two surface meshes avoiding voxelization errors of voxel-based comparison, according to \cref{eq:Dice}. The visualization utilities \texttt{plot\_mesh} and \texttt{plot\_sphere\_pack} provide interactive three-dimensional views based on \texttt{pyvista}.

\section{Illustrative examples}

\begin{figure}
    \centering
    \includegraphics[width=1.0\linewidth]{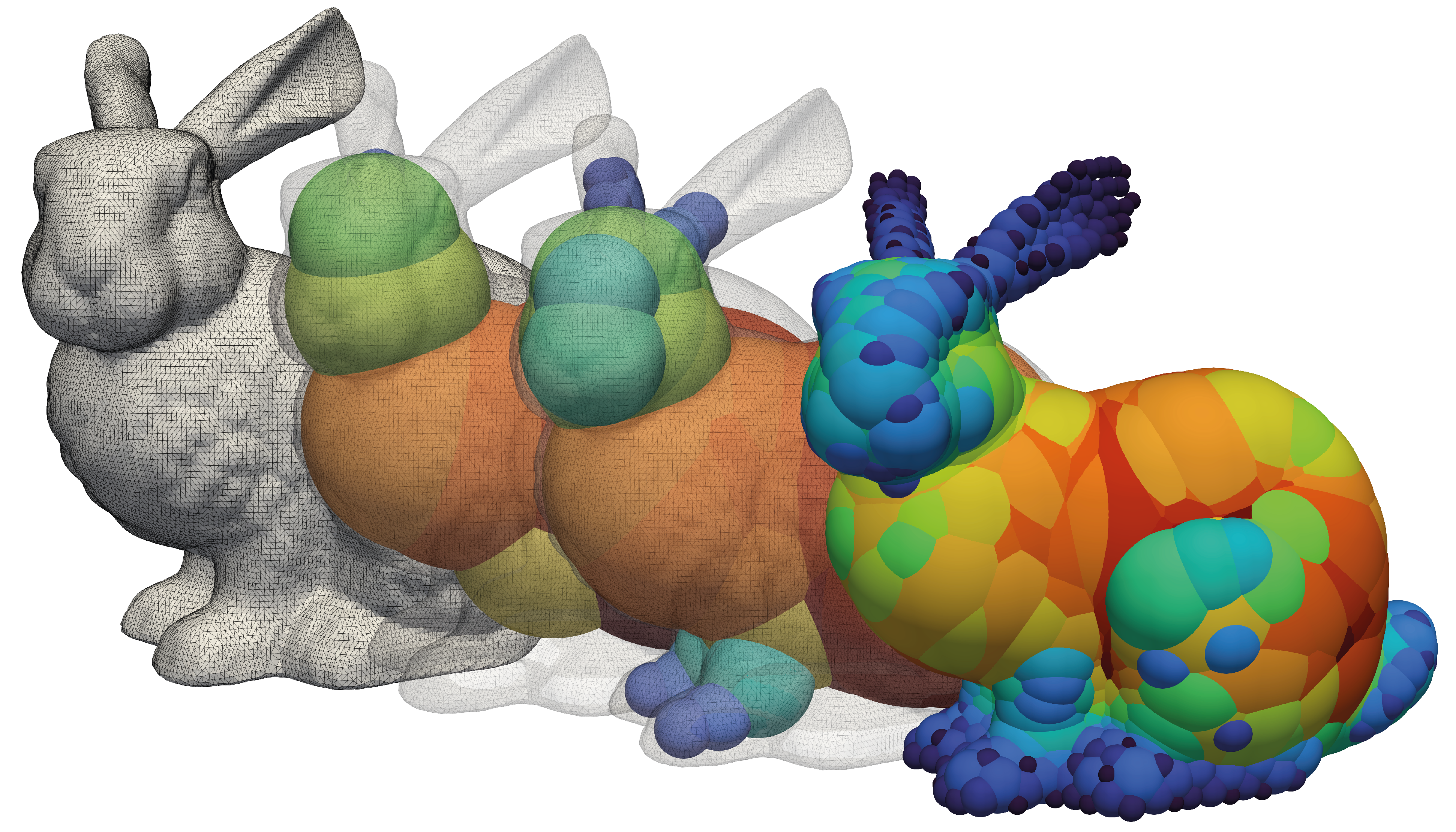}
    \caption{Surface mesh of the Stanford Bunny \cite{StanfordBunny} and its multisphere reconstruction with 10, 50, and 629 spheres.}
    \label{fig:stanford_bunny}
\end{figure}

\Cref{fig:stanford_bunny} shows multi-sphere models of the Stanford Bunny \cite{StanfordBunny} reconstructed with $n=\{10, 50, 629\}$ spheres. The corresponding Python script is shown in \Cref{fig:code_example}. 

\begin{figure}[]
  \centering
  \definecolor{codebg}{rgb}{0.95,0.95,0.95}
  \lstset{
  basicstyle=\ttfamily\small,
  backgroundcolor=\color{codebg},
  breaklines=true,
  columns=fullflexible,
  keepspaces=true,
  frame=single,
  showstringspaces=false,
  numbers=left,
  numberstyle=\tiny,
  numbersep=8pt
}

  \begin{lstlisting}[style=mypython]
import multisphere as mss

# multi-sphere reconstruction
mesh = mss.load_mesh(stanford_bunny.stl)
sphere_pack = mss.multisphere_from_mesh(
    mesh=mesh,
    div=150,
    min_radius_vox=8
    precision=0.95,
    min_center_distance_vox=4,
    max_spheres=10,
    confine_mesh=True)

# compute mesh-based dice coefficient
sphere_mesh = mss.create_multisphere_mesh(
    sphere_pack)

dice = mss.compute_dice_coefficient(
    mesh_1=mesh,
    mesh_2=sphere_mesh)

# visualization and export
mss.plot_sphere_pack(sphere_pack)
mss.export_sphere_pack_to_csv(
sphere_pack, path)
mss.export_sphere_pack_to_vtk(
sphere_pack, path)
\end{lstlisting}
    \caption{\texttt{multisphere} script to create the multisphere model of the Stanford Bunny shown in \cref{fig:stanford_bunny}.}
    \label{fig:code_example}
\end{figure}

\begin{figure}
    \centering
    \includegraphics[width=1.0\linewidth]{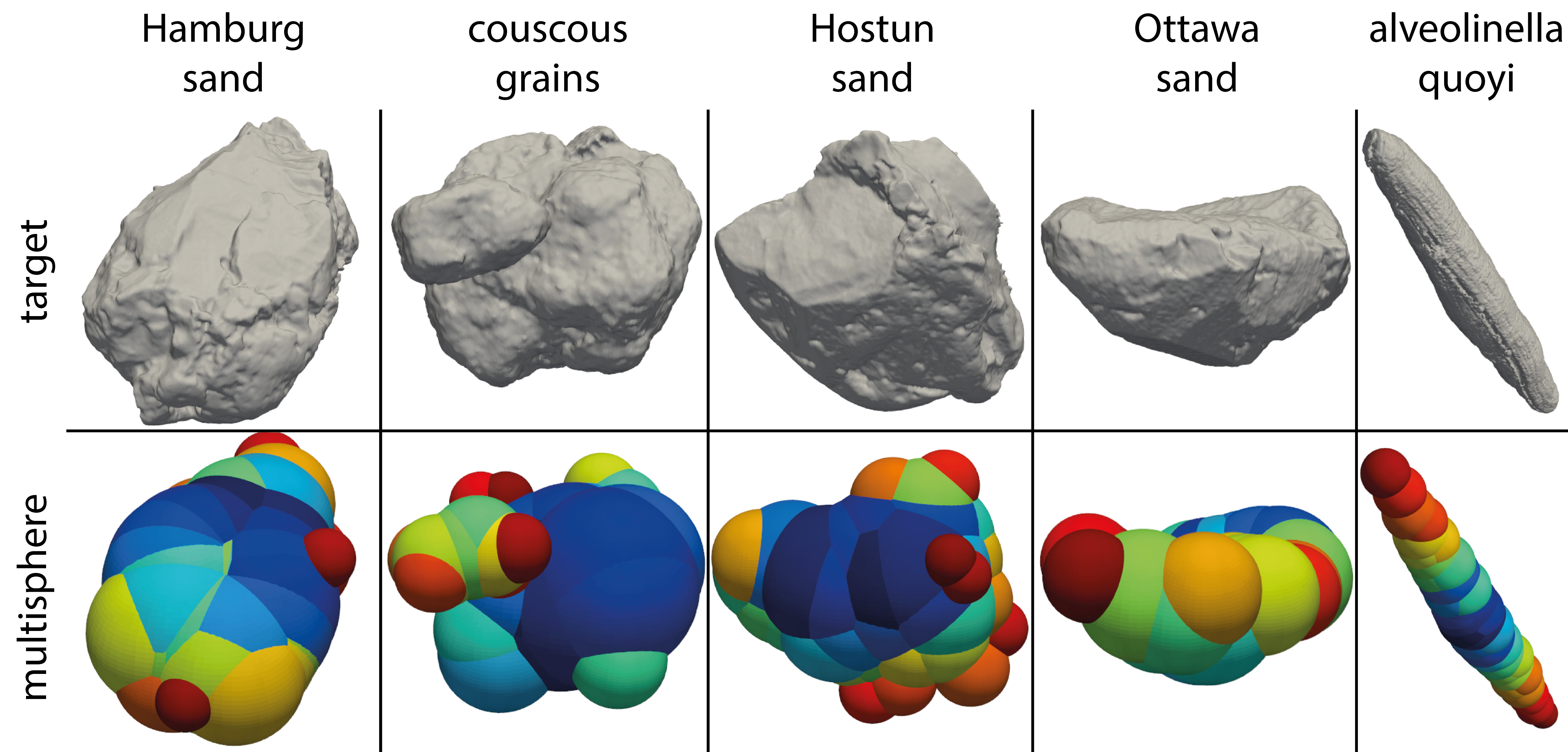}
    \caption{Sample grains from the Sand Atlas \cite{Milatz.2021}. Top row: surface meshes of Hamburg sand \cite{Milatz.2021}, couscous grains \cite{Vego.2023}, Hostun sand \cite{Wiebicke.2017}, Ottawa sand \cite{Saadatfar.2012}, and Alveolinella quoyi \cite{Luijmes.2024}. Bottom row: corresponding multi-sphere models generated with \texttt{multisphere}.
    }
    \label{fig:SAND-examples}
\end{figure}

\Cref{fig:SAND-examples} shows sample grains of five granular materials taken from the Sand Atlas and corresponding multisphere-models. These examples cover a broad range of particle morphologies, including angular, rounded, elongated, and flat shapes. 

\begin{figure}
    \centering
    \includegraphics[width=1.0\linewidth]{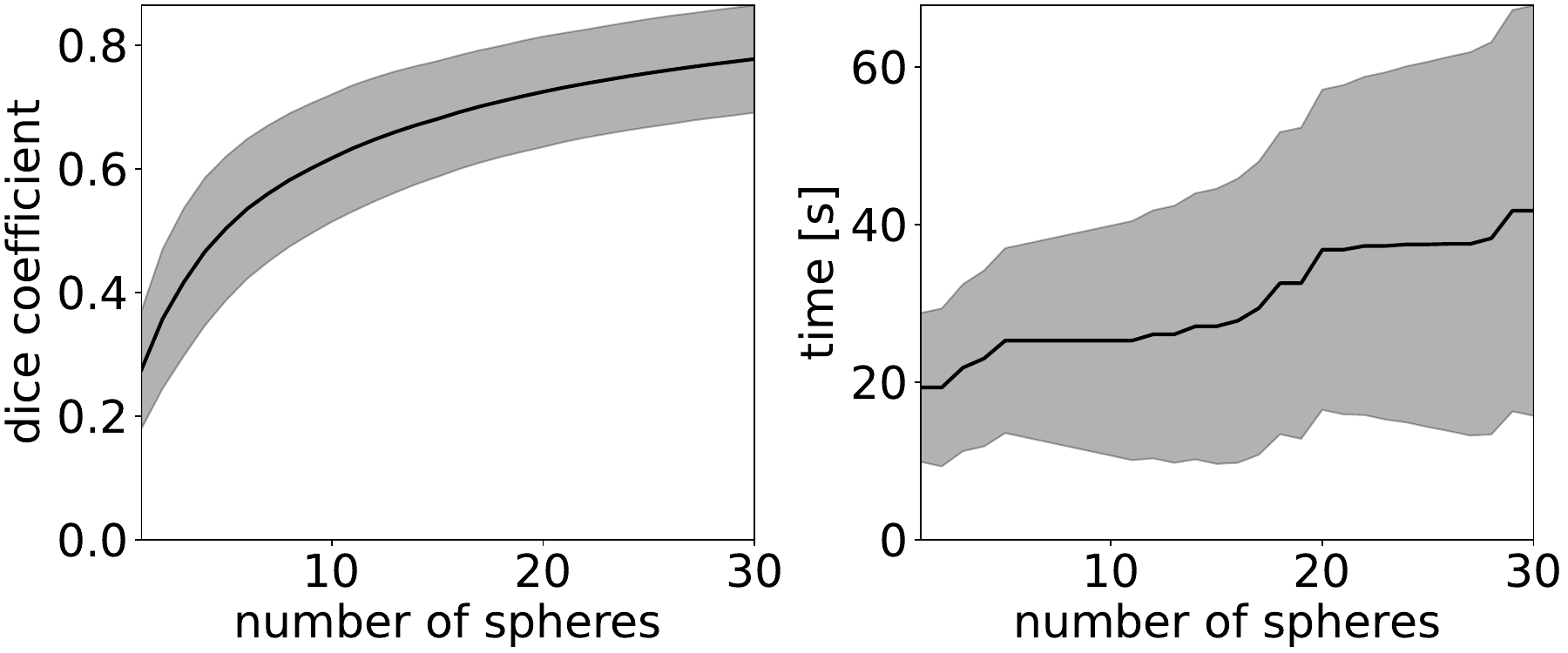}
    \caption{Dice coefficient and \texttt{multisphere} runtime for the reconstruction of all 402 particles from the Sand Atlas \cite{Milatz.2021} as a function of the number of spheres.
    }
    \label{fig:sand_atlas_fidelity}
\end{figure}

\Cref{fig:sand_atlas_fidelity} (left) shows the Dice coefficient, \cref{eq:Dice}, for the reconstruction of multi-sphere models for all granular particles from the Sand Atlas \cite{Milatz.2021} as a function of the number $n$ of spheres. The solid line represents the mean value $\left<D\right>$ over all grains, and the shaded region indicates the standard deviation.

\Cref{fig:sand_atlas_fidelity} (right) shows the computer time (arbitrary units, depending on the hardware). A comparative evaluation against other multi-sphere reconstruction algorithms is presented in \cite{Buchele.2026}.

\section{Impact and Conclusions}

\texttt{multisphere} is an open-source implementation of the MSS algorithm in Python. The software generates multisphere models for use in DEM simulations based on voxelised input data, as well as surface meshes, which allows for the direct use of target particle shapes from tomography, 3D scans or CAD.

By generating multi-sphere representations with fewer spheres at a prescribed reconstruction accuracy \cite{Buchele.2026}, \texttt{multisphere} reduces the computational cost of shape-resolved DEM simulations, where the number of sphere--sphere interactions is a key performance factor. This is particularly relevant for simulations involving large numbers of irregular particles and for simulations in which particle shapes must be reconstructed repeatedly during runtime, such as fragmentation and other plastic deformation of the DEM particles. 





\bibliographystyle{elsarticle-num}
\bibliography{softwareX_references}

@article{Poeschel.1993,
  title = {Static friction phenomena in granular materials: {C}oulomb law versus particle geometry},
  author = {P\"oschel, Thorsten and Buchholtz, Volkhard},
  journal = {Physical Review Letters},
  volume = {71},
  pages = {3963--3966},
  year = {1993},
  doi = {10.1103/PhysRevLett.71.3963}
}

@article{Buchholtz.1994,
author = {Buchholtz, Volkhard and Pöschel, Thorsten},
 doi = {10.1016/0378-4371(94)90467-7},
 journal = {Physica A-Statistical Mechanics and Its Applications},
 pages = {390-401},
 title = {Numerical investigations of the evolution of sandpiles},
 volume = {202},
 year = {1994},
}

@article{Buchholtz.1996,
    author = {Buchholtz, V. and P\"oschel, T.},
    title = {Avalanche statistics of sand heaps},
    journal = {Journal of Statistical Physics},
volume = {84},
pages = {1373–1378},
    year = {1996},
doi = {10.1007/BF02174136},
}

@article{Hubbard.1996, 
author = {Hubbard, Philip M.}, 
title = {Approximating polyhedra with spheres for time-critical collision detection}, 
year = {1996}, 
volume = {15}, 
doi = {10.1145/231731.231732}, 
journal = {ACM Transactions on Graphics}, 
pages = {179–210}, 
}

@article{Favier.1999,
author = {Favier, John and Abbaspour-Fard, Mohammad and Kremmer, M. and Raji, A. O.},
year = {1999},
pages = {467-480},
title = {Shape representation of axi-symmetrical, non-spherical particles in discrete element simulation using multi-element model particles},
volume = {16},
journal = {Engineering Computations},
doi = {10.1108/02644409910271894}
}

@article{Angelidakis.2021,
title = {{CLUMP}: {A} Code Library to generate Universal Multi-sphere Particles},
journal = {SoftwareX},
volume = {15},
pages = {100735},
year = {2021},
doi = {10.1016/j.softx.2021.100735},
author = {Angelidakis, Vasileios and  Nadimi, Sadegh and Otsubo, Masahide and Utili, Stefano}
}

@article{Canbolat.2025,
title = {A {P}ython implementation of {CLUMP}, the Code Library to generate Universal Multi-sphere Particles},
journal = {SoftwareX},
volume = {29},
pages = {101957},
year = {2025},
doi = {10.1016/j.softx.2024.101957},
author = {Canbolat, Ahmet Utku and Nadimi, Sadegh  and Angelidakis, Vasileios}
}

@inproceedings{Amberger.2012,
author = {Amberger, Stefan and Friedl, Michael and Goniva, Christoph and Pirker, Stefan and Kloss, Christoph},
year = {2012},
pages = {},
title = {Approximation of Objects by Spheres for Multisphere Simulations in {DEM}},
journal = {ECCOMAS 2012 - European Congress on Computational Methods in Applied Sciences and Engineering},
}

@article{Ferellec.2010,
  title={A method to model realistic particle shape and inertia in {DEM}},
  author={Ferellec, Jean-Francois and McDowell, Glenn R},
  journal={Granular Matter},
  volume={12},
  pages={459--467},
  year={2010},
  doi={10.1007/s10035-010-0205-8}
}

@Article{Milatz.2021,
author={Milatz, Marius
and H{\"u}sener, Nicole
and And{\`o}, Edward
and Viggiani, Gioacchino
and Grabe, J{\"u}rgen},
title={Quantitative {3D} imaging of partially saturated granular materials under uniaxial compression},
journal={Acta Geotechnica},
year={2021},
volume={16},
pages={3573-3600},
doi={10.1007/s11440-021-01315-5}
}

@phdthesis{Vego.2023,
author = {Vego, Ilija},
title = {Multi-modal investigation of hygroscopic granular media at high relative humidity},
school = {Université Grenoble Alpes},
year = {2023}
}

@ARTICLE{Wiebicke.2017,
title     = "On the metrology of interparticle contacts in sand from x-ray tomography images",
author    = "Wiebicke, Max and And{\`o}, Edward and Herle, Ivo and Viggiani, Gioacchino",
journal   = "Meas. Sci. Technol.",
publisher = "IOP Publishing",
volume    =  28,
number    =  12,
pages     = "124007",
month     =  dec,
year      =  2017,
doi={10.1088/1361-6501/aa8dbf}
}

@INPROCEEDINGS{Saadatfar.2012,
title      = "{3D} mapping of deformation in an unconsolidated sand: {A} micro mechanical study",
booktitle  = "{SEG} Technical Program Expanded Abstracts 2012",
author     = "Saadatfar, Mohammad and Francois, Nicolas and Arad, Alon and Madadi, Mahyar and Cruikshank, Ron and Alizadeh, Mehdi and Sheppard, Adrian and Kingston, Andrew and Limay, Ajay and Senden, Tim and Knackstedt, Mark",
publisher  = "Society of Exploration Geophysicists",
month      =  sep,
year       =  2012,
pages={1-6},
doi={10.1190/segam2012-1263.1}
}

@ARTICLE{Luijmes.2024,
author={Luijmes, Joost
and van Leeuwen, Tristan
and Renema, Willem},
title={ForametCeTera, a novel CT scan dataset to expedite classification research of (non-)foraminifera},
journal={Scientific Data},
year={2024},
volume={11},
pages={642},
doi={10.1038/s41597-024-03476-w}
}

@article{Dice.1945,
author = {Dice, Lee R.},
title = {Measures of the Amount of Ecologic Association Between Species},
journal = {Ecology},
volume = {26},
number = {3},
pages = {297-302},
doi = {10.2307/1932409},
year = {1945}
}

@misc{Cow2:Viewpoint,
  title        = {cow2 [3D model] from the Princeton Suggestive Contour Gallery},
  author       = {{Viewpoint Animation Engineering} and {Sun Microsystems}},
  howpublished = {\url{https://gfx.cs.princeton.edu/proj/sugcon/models/}},
  note         = {Accessed: 2026-01-26},
  year         = {n.d.},
}

@book{Matuttis.2014,
author = {Matuttis, Hans-Georg and Chen, Jian},
title = {Understanding the Discrete Element Method: Simulation of Non-Spherical Particles for Granular and Multi-body Systems},
publisher = {Wiley},
doi={10.1002/9781118567210},
year = {2014},
}

@article{Cundall.1979,
author = {Cundall, P. A. and Strack, O. D. L.},
title = {A discrete numerical model for granular assemblies},
doi={10.1680/geot.1979.29.1.47},
journal = {G\'eotechnique},
volume = {79},
pages = {47-65},
year = {1979},
}

@book{Poeschel.2005,
author    = {Thorsten Pöschel and Thomas Schwager},
title     = {Computational Granular Dynamics: Models and Algorithms},
publisher = {Springer},
address   = {Berlin, Heidelberg},
year      = {2005},
doi       = {10.1007/3-540-27720-X}
}

@article{Lu.2015,
title = {Discrete element models for non-spherical particle systems: From theoretical developments to applications},
journal = {Chemical Engineering Science},
volume = {127},
pages = {425-465},
year = {2015},
issn = {0009-2509},
doi = {10.1016/j.ces.2014.11.050},
author = {G. Lu and J.R. Third and C.R. Müller}
}

@article{Zhao.2023,
title = {The role of particle shape in computational modelling of granular matter},
journal = {Nature Reviews Physics},
volume = {5},
year = {2023},
doi = {10.1038/s42254-023-00617-9},
author = {Jidong Zhao and Shiwei Zhao and Stefan Luding}
}

@article{Buchele.2026,
title = {Multi-sphere shape generator for DEM simulations of complex-shaped particles},
journal = {Powder Technology},
year = {2026},
author = {Buchele Felix and Thorsten Pöschel and Patric Müller},
note = {, submitted},
doi = {10.48550/arXiv.2603.05877}
}

@Article{Levy.2025,
author={Levy, Avivit
and Shalom, B. Riva
and Chalamish, Michal},
title={A guide to similarity measures and their data science applications},
journal={Journal of Big Data},
year={2025},
month={Jul},
day={26},
volume={12},
number={1},
pages={188},
issn={2196-1115},
doi={10.1186/s40537-025-01227-1}
}

@misc{StanfordBunny,
author={{Stanford Computer Graphics Laboratory}},
title={The Stanford 3D Scanning Repository},
howpublished={\url{http://graphics.stanford.edu/data/3Dscanrep/}},
note={{S}tanford {B}unny {M}odel, accessed: 2026-03-19},
year={1994}
}

\end{document}